\documentclass[reprint,nofootinbib,amsmath,amssymb,aps,prd]{revtex4-1}

\usepackage{graphicx}
\usepackage{dcolumn}
\usepackage{bm}
\usepackage{microtype}
\numberwithin{equation}{section}
\usepackage{hyperref}
\usepackage{xcolor}

\begin{document}

\title{Bouncing cosmology in $f(R,\mathcal{G})$ gravity by order reduction}

\author{Bruno J. Barros}
\author{Elsa M. Teixeira}
\author{Daniele Vernieri}
\affiliation{Instituto de Astrof\'isica e Ci\^encias do Espa\c{c}o, 
Faculdade de Ci\^encias da Universidade de Lisboa, Campo Grande, PT1749-016 Lisboa, Portugal}
\date{\today}

\begin{abstract}
A bouncing universe is a viable candidate to solve the initial singularity problem. Here we consider bouncing solutions in the context of $f(R,\mathcal{G})$ gravity by using an order reduction technique which allows one to find solutions that are perturbatively close to General Relativity. This procedure also acts as a model selection approach. Indeed, several covariant gravitational actions leading to a bounce are directly selected by demanding that the Friedmann equation derived within such gravity theories coincides with the one emerging from Loop Quantum Cosmology.
\end{abstract}

\maketitle
\onecolumngrid

\section{Introduction}

An effective description of Loop Quantum Cosmology (LQC)~\cite{Bojowald:2006da,Ashtekar:2011ni} enables one to write cosmological dynamics in a classical fashion incorporating leading order quantum corrections. While at low energies the predictions of LQC are very close to General Relativity (GR), in high density regimes, where quantum effects become dominant, these might drastically depart from the classical description~\cite{Tsujikawa:2003vr,Bojowald:2004xq,Singh:2003au}. One such phenomenon is the Big Bang singularity, which in light of LQC is replaced by a quantum bounce~\cite{Bojowald:2001xe,Bojowald:2004kt,Ashtekar:2006wn,Brandenberger:2012zb,Odintsov:2014gea,Odintsov:2015ynk,Odintsov:2016tar}. This becomes evident when writing the modified Friedmann equation emerging from the effective description of LQC~\cite{Taveras:2008ke,Banerjee:2011qu}:
\begin{equation}
\label{LQCFriedmann}
H^2 = \frac{\kappa}{3}\rho \left( 1-\frac{\rho}{\rho_c} \right),
\end{equation}
where $H := \dot{a}/a$ is the Hubble rate, $a(t)$ is the scale factor of the universe, $\kappa \equiv 8\pi$, and $\rho_c = \sqrt{3}/(32\pi^2\gamma^3)$ gives the critical energy density for the bounce, with $\gamma$ being the Barbero--Immirzi parameter. Notice that we are using physical units such that the Newton's gravitational constant $G_N$, the speed of light $c$ and the reduced Planck constant $\hslash$ have been set to one.

We will consider bouncing scenarios in the context of modified gravity theories~\cite{Capozziello:2007ec,Nojiri:2017ncd} whose gravitational action contains higher-order derivatives and that generically propagate extra degrees of freedom. In the specific case of the generalized Gauss--Bonnet $f(R,\mathcal{G})$ gravity theories, that will be the main focus of this work, $R$ being the Ricci scalar and $\mathcal{G}$ the Gauss--Bonnet term, one can indeed write equivalently the corresponding higher-order gravitational action as a scalar--tensor theory with two extra dynamical scalar fields~\cite{DeFelice:2009ak}. 

These generalized Gauss--Bonnet gravity models have been considered to describe cosmological phenomena, such as the inflationary epoch~\cite{Odintsov:2018nch,Odintsov:2020sqy} and dark energy scenarios~\cite{Alimohammadi:2008fq,Odintsov:2018nch,Elizalde:2010jx}. Future singularities within $f(R,\mathcal{G})$ gravity have been discussed in Ref.~\cite{Bamba:2009uf}, together with some proposals to avoid them, such as considering a Lagrangian with terms of the form $\mathcal{G}^{\alpha}/R^{\beta}$. The study of the stability conditions for scalar and tensor perturbations in $f(R,\mathcal{G})$ gravity was conducted in Refs.~\cite{DeFelice:2009ak,DeFelice:2010sh,DeFelice:2010hg}. Primordial inflation was also studied in Ref.~\cite{DeLaurentis:2015fea}, where it was shown that a double inflationary setting naturally arises in the context of such theories. Bouncing scenarios and cosmological applications of $f(\mathcal{G})$ models have been studied in Refs.~\cite{Oikonomou:2015qha,Li:2007jm}.

In order to avoid the presence of spurious degrees of freedom that are usually present in such theories, we will implement the so-called order reduction technique~\cite{Bel:1985zz,Simon:1990ic,Simon:1991bm}. The latter consists in reducing the field equations by expressing the geometrical quantities in terms of the matter fields at the lowest order in the perturbative expansion. Hence, the gravity theory at hand is framed within an effective field theory approach and the reduced field equations contain derivatives up to second order. Then, only the solutions that are perturbatively close to GR are selected.

This paper addresses the following questions: Can the Friedmann equation emerging from LQC be derived from an effective modified gravity theory? And, if so, what is the most general form for the Lagrangian under these conditions? Indeed, this was already proved to be possible for $f(R)$~\cite{Sotiriou:2008ya} and $f(\mathcal{G})$~\cite{Terrucha:2019jpm} gravity. In the present manuscript we approach this issue in the context of $f(R,\mathcal{G})$ gravity models. We find novel solutions for the form of the gravitational Lagrangian allowing for bouncing cosmological scenarios by means of the order reduction technique. Moreover this work extends the results reported in Refs.~\cite{Sotiriou:2008ya, Terrucha:2019jpm}. Notice that previous works have been conducted to reproduce the Friedmann equation from LQC but using different methods than the one employed in this work and within distinct gravity theories. In particular, in the framework of Palatini $f(R)$ gravity~\cite{Olmo:2008nf}, by demanding that the Friedmann equation matches the one from LQC, the specific form of the $f(R)$ Lagrangian was numerically reconstructed by means of interpolation techniques. Moreover, the covariant action for a class of higher-derivative scalar--tensor theories, referred to as mimetic gravity, was reconstructed in Ref.~\cite{Liu:2017puc} such that the effective dynamics of LQC is reproduced within homogeneous and isotropic backgrounds, resorting to an Hamiltonian analysis. 

The structure of the paper is the following. In Sec.~\ref{model} we write the full action for the gravitational sector and for the matter fields, the modified field equations, and then apply the order reduction technique. In Sec.~\ref{sectionFLRW} we follow to specify the background cosmology and write the evolution equations accordingly. The results are presented and discussed in Sec.~\ref{bouncing}. Finally we conclude in Sec.~\ref{conclusions}.

\section{$f(R,\mathcal{G})$ gravity and order reduction}
\label{model}

We begin by writing down the action for $f(R,\mathcal{G})$ modified gravity theories in a 4-dimensional spacetime,
\begin{equation}
\label{action}
\mathcal{S}= \frac{1}{2\kappa}\int d^4 x \sqrt{-g}\big[ R + f(R,\mathcal{G})  \big] + \mathcal{S}_m (g_{\mu\nu},\psi)\,,
\end{equation}
where $g$ is the determinant of the metric tensor and the gravitational sector consists of the standard Einstein--Hilbert term plus an additional general function of the Ricci scalar curvature and the Gauss--Bonnet term, the latter defined as
\begin{equation}
\mathcal{G} = R^2 - 4R^{\mu\nu}R_{\mu\nu} + R^{\mu\nu\alpha\beta}R_{\mu\nu\alpha\beta}\,,
\end{equation}
with $R_{\mu \nu}$ and $R_{\mu \nu \alpha \beta}$ being the components of the Ricci and Riemann tensors, respectively. The deviations from GR are generically enclosed in the function $f(R,\mathcal{G})$. The object $\mathcal{S}_m$ in Eq.~\eqref{action} portrays the matter sector, with $\psi$ standing for a collective representation of the matter fields. 

Following the standard procedure of varying the action in Eq.~\eqref{action} with respect to (wrt) the metric $g^{\mu\nu}$, we arrive at the following modified field equations:
\begin{gather}
R_{\mu\nu} - \frac{1}{2}g_{\mu\nu}R - \frac{1}{2}g_{\mu\nu} f + f_{,R}R_{\mu\nu}+ \left( g_{\mu\nu}\,\square  - \nabla_{\mu}\nabla_{\nu} \right)f_{,R} 
+ f_{,\mathcal{G}} \left( 2RR_{\mu\nu} - 4R_{\mu}^{\,\,\,\,\alpha}R_{\nu\alpha} + 2R_{\mu\alpha\beta\gamma}R_{\nu}^{\,\,\,\,\alpha\beta\gamma}  -4R_{\mu\alpha\nu\beta}R^{\alpha\beta} \right) \nonumber \\
+ \left(2\,g_{\mu\nu}R\,\square -2R\, \nabla_{\mu}\nabla_{\nu} +4R_{\mu}^{\,\,\,\,\alpha}\,\nabla_{\alpha}\nabla_{\nu}+4R_{\nu}^{\,\,\,\,\alpha}\,\nabla_{\alpha}\nabla_{\mu} - 4g_{\mu\nu}R^{\alpha\beta}\,\nabla_{\alpha}\nabla_{\beta} - 4R_{\mu\nu}\,\square + 4R_{\mu\,\,\,\,\nu}^{\,\,\,\,\alpha\,\,\,\,\beta}\,\nabla_{\alpha}\nabla_{\beta} \right)f_{,\mathcal{G}}
=\kappa T_{\mu\nu}\,, \label{fieldEqs}
\end{gather}
where $\nabla_{\mu}$ is the covariant derivative and $\square = g^{\mu \nu} \nabla_{\mu} \nabla_{\nu}$ is the D'Alembertian operator. We have defined $f_{,R} := \partial f/ \partial R$, $f_{,\mathcal{G}} :=  \partial f/ \partial \mathcal{G}$, and
\begin{equation}
T_{\mu\nu}= -\frac{2}{\sqrt{-g}}\frac{\delta \mathcal{S}_m}{\delta g^{\mu\nu}}\,
\end{equation}
is the matter stress-energy tensor.

We rely on the fact that the Riemann tensor can be decomposed into a traceless, semi-traceless and a scalar part, respectively as:
\begin{eqnarray}
R_{\mu\nu\alpha\beta} = &&\, C_{\mu\nu\alpha\beta}  + \frac{1}{2}\left( g_{\mu\alpha}R_{\beta\nu}+g_{\nu\beta}R_{\alpha\mu}- g_{\mu\beta}R_{\alpha\nu}  -g_{\nu\alpha}R_{\beta\mu} \right) +\frac{1}{6}\left( g_{\mu\beta}g_{\alpha\nu} - g_{\mu\alpha}g_{\beta\nu} \right) R\,,
\end{eqnarray}
with $C_{\mu\nu\alpha\beta}$ being the Weyl tensor components. As a first simplification, we require that the Weyl tensor vanishes: $C_{\mu\nu\alpha\beta}=0$. This amounts to assuming that our spacetime is conformally flat, {\it i.e.}, every point in our spacetime manifold contains a neighborhood which can be mapped to a flat space by means of a conformal transformation. This condition always holds for a Friedmann--Lema\^itre--Robertson--Walker (FLRW) background on which we will base our study throughout the next sections. Thus, the Riemann tensor components reduce to
\begin{eqnarray}
R_{\mu\nu\alpha\beta} = &&\, \frac{1}{2}\left( g_{\mu\alpha}R_{\beta\nu}+g_{\nu\beta}R_{\alpha\mu} - g_{\mu\beta}R_{\alpha\nu} -g_{\nu\alpha}R_{\beta\mu} \right) +\frac{1}{6}\left( g_{\mu\beta}g_{\alpha\nu} - g_{\mu\alpha}g_{\beta\nu} \right) R\,,
\label{riemtens}
\end{eqnarray}
and the Gauss--Bonnet term $\mathcal{G}$ can be written as
\begin{equation}
\mathcal{G} = \frac{2}{3}R^2 - 2R_{\mu\nu}R^{\mu\nu}\,.
\label{gaussbon}
\end{equation}
In order to avoid the presence of spurious degrees of freedom that usually arise in these theories, we employ the order reduction technique~\cite{Bel:1985zz,Simon:1990ic,Simon:1991bm} by introducing a parameter $\epsilon$ such that its vanishing promptly leads us back to GR. Thus, following the procedure presented in Refs.~\cite{Sotiriou:2008ya,Terrucha:2019jpm}, we parametrize the function $f(R,\mathcal{G})$ as
\begin{equation}
\label{fparam}
f(R,\mathcal{G}) = 2\Lambda + \epsilon \varphi(R,\mathcal{G})\,,
\end{equation}
where $\Lambda$ is a cosmological constant term and the dimensionless parameter $\epsilon$ regulates the deviations from GR. This is consistent with the fact that we wish to frame $f(R,\mathcal{G})$ gravity within an effective field theory approach such that its solutions are perturbatively close to GR, namely that $\epsilon\varphi\ll R$. Accordingly, we will only consider the lowest order terms in $\epsilon$. By doing so, we constrain the higher order field equations to contain up to second order derivatives of the metric, thus avoiding the propagation of additional degrees of freedom arising in such theories.

To apply the order reduction technique we simply replace all the $\mathcal{O}(\epsilon)$ terms in the modified field equations by their lowest order ($\epsilon =0$) counterpart, that ends up with expressing the geometrical quantities $R$, $R_{\mu\nu}$, $R_{\mu\nu\alpha\beta}$, and $\mathcal{G}$, in terms of the matter fields at the lowest order in the perturbative expansion. Indeed, evaluating Eq.~\eqref{fieldEqs} and its trace, with $f$ as defined in Eq.~\eqref{fparam}, for $\epsilon=0$, yields the following expressions for the order reduced curvature scalar and Ricci tensor components:
\begin{eqnarray}
R^T &=& -\kappa T -4\Lambda\,, \label{orr2} \\ 
R^T_{\mu\nu} &=& \kappa T_{\mu\nu}-\frac{\kappa}{2}g_{\mu\nu}T - \Lambda g_{\mu\nu}\,,
\label{orr}
\end{eqnarray}
where the superscript $T$ denotes the order reduced quantities. Based on the above relations, together with the assumption of vanishing Weyl tensor, the Riemann tensor components in Eq.~\eqref{riemtens}, and the Gauss--Bonnet invariant in Eq.~\eqref{gaussbon}, can be expressed at the lowest order as follows:
\begin{eqnarray}
R^T_{\mu\nu\alpha\beta} = &&\,\,-\frac{\kappa}{2}\left( g_{\mu\beta}T_{\alpha\nu}+g_{\nu\alpha}T_{\beta\mu} - g_{\mu\alpha}T_{\beta\nu}-g_{\nu\beta}T_{\alpha\mu} \right) -\frac{1}{3}\left( g_{\mu\alpha}g_{\nu\beta}-g_{\mu\beta}g_{\alpha\nu}\right) \left( \Lambda+\kappa T \right), \label{orr3}\\
\mathcal{G}^T = &&\,\, \frac{2}{3}\kappa^2 T^2 - 2 \kappa^2 T_{\mu\nu}T^{\mu\nu} + \frac{8}{3}\Lambda^2 + \frac{4}{3}\Lambda\kappa T\,. \label{orr4}
\end{eqnarray}
Finally, we can formally write the reduced field equations, with each order-$\epsilon$ term replaced by its corresponding order reduced analogue,
\begin{gather}
R_{\mu\nu} - \frac{1}{2}g_{\mu\nu}R- g_{\mu\nu} \Lambda +\epsilon\bigg[ -\frac{1}{2}g_{\mu\nu} \varphi^T + \varphi^T_{,R}R^T_{\mu\nu}+ \left( g_{\mu\nu}\,\square  - \nabla_{\mu}\nabla_{\nu} \right)\varphi^T_{,R} 
+ \varphi^T_{,\mathcal{G}} \left( 2R^T R^T_{\mu\nu} - 4R_{\mu}^{T\,\,\,\alpha}R^T_{\nu\alpha}  \right.  \nonumber \\
\left. + 2R^T_{\mu\alpha\beta\gamma}R_{\nu}^{T\,\,\,\alpha\beta\gamma} -4R^T_{\mu\alpha\nu\beta}R^{T\,\,\alpha\beta} \right)+ \left(2\,g_{\mu\nu}R^T\,\square -2R^T\, \nabla_{\mu}\nabla_{\nu} +4R_{\mu}^{T\,\,\alpha}\,\nabla_{\alpha}\nabla_{\nu} +4R_{\nu}^{T\,\,\alpha}\,\nabla_{\alpha}\nabla_{\mu} - 4g_{\mu\nu}R^{T\,\,\alpha\beta}\,\nabla_{\alpha}\nabla_{\beta}  \right. \nonumber \\
\left. - 4R^T_{\mu\nu}\,\square + 4R_{\mu\,\,\,\,\nu}^{T\alpha\,\,\,\,\beta}\,\nabla_{\alpha}\nabla_{\beta} \right)\varphi_{,\mathcal{G}}\bigg] =\kappa T_{\mu\nu}\,, \label{ORfield}
\end{gather}
in which, accordingly, the derivatives are wrt the lowest order quantities, {\it i.e.}, $\varphi^T_{,R} := \partial \varphi^T / \partial R^T$, with $\varphi^T := \varphi^T (R^T,\mathcal{G}^T)$. 

Equation~\eqref{ORfield} is then simply written as GR plus order-$\epsilon$ terms which will dominate near the bounce. The assumption $\epsilon\varphi\ll R$ is a necessary condition for the whole scheme to be valid~\cite{Sotiriou:2008ya,Terrucha:2019jpm}, guaranteeing that the corrective terms are treated within an effective field theory approach, where $R\ll \rho_c \sim l_p^{-2}$ ($l_p$ being the Planck scale) as will be verified {\it a posteriori} through the bouncing solutions presented in Sec.~\ref{bouncing}.

We proceed with the implementation of this model to a particular geometrical setting, specifying a form for $g_{\mu\nu}$ and writing the cosmological equations in this context.

\section{Order reduction within a FLRW geometry}
\label{sectionFLRW}

We will now consider the Friedmann--Lema\^itre--Robertson--Walker (FLRW) cosmological background, with line element given by
\begin{equation}
\label{frwmetric}
ds^2 = -dt^2 +a(t)^2 \left[ \frac{dr^2}{1-kr^2} +r^2\left( d\theta^2 + \sin^2\theta\, d\phi^2 \right) \right].
\end{equation}
The constant $k\in\left\{-1,0,1\right\}$ determines the spatial curvature of the universe (hyperbolic, flat or spherical, respectively).

Furthermore, we assume a perfect fluid form for the matter stress-energy tensor, namely
\begin{equation}\label{stress}
T_{\mu\nu} = \left( \rho + p \right)u_{\mu}u_{\nu} + p\,g_{\mu\nu}\,,
\end{equation}
where $u_{\mu}$ is the fluid's 4-velocity, defined in such a way that $u_{\mu} u^{\mu}=-1$, and $\rho$ and $p$ denote the energy density and pressure, respectively. We consider a barotropic equation of state (EoS) for the pressure, {\it i.e.},
\begin{equation}\label{press}
p = w\rho\,,
\end{equation}
with $w\in[-1,1]$ being a constant.

The time component of the conservation equation, $\nabla_{\mu}T^{\mu\nu}=0$, yields the continuity equation
\begin{equation}
\label{continuity}
\dot{\rho}= -3H \left( 1+w \right)\rho,
\end{equation}
where a dot denotes derivative wrt cosmic time.

Using Eqs.~\eqref{stress}-\eqref{press}, the order reduced scalars $R^T$ and $\mathcal{G}^T$ read:
\begin{eqnarray}
R^T &=&  \kappa\rho\left( 1-3w \right) -4\Lambda\,,  \label{ricciOR} \\
\mathcal{G}^T &=& \frac{4}{3}\left( \Lambda - \kappa\rho \right) \big[ 2\Lambda + \kappa\rho\left( 1+3w \right) \big]. \label{GBOR}
\end{eqnarray}

The modified Friedmann equation is computed from the $0$--$0$ component of the reduced field equations in Eq.~\eqref{ORfield}:
\begin{gather}
6H^2 + \frac{6k}{a^2} -2\Lambda+2\kappa\rho+ \epsilon  \bigg\{ \varphi^T +6 H \dot{\varphi}^T_R +\varphi^T_R \Big[2 \Lambda +\kappa\rho (1 +3  w) \Big]-8 H \dot{\varphi}^T_{\mathcal{G}} ( \Lambda -\kappa \rho) \nonumber \\
 -\frac{4}{3}   \varphi^T_{\mathcal{G}} (\Lambda -\kappa  \rho) \Big[2 \Lambda +\kappa\rho (1 +3  w)\Big]\bigg\} =0\,.
 \label{fried}
\end{gather}

It is convenient to use the chain rule, alongside the continuity equation, Eq.~\eqref{continuity}, and Eqs.~\eqref{ricciOR}-\eqref{GBOR}, to express the time derivative of $\varphi^T$ in terms of derivatives wrt $\mathcal{G}^T$ and $R^T$: 
\begin{equation}
\dot{\varphi}^T = \frac{\partial \varphi^T}{\partial R^T}\frac{\partial R^T}{\partial \rho}\dot{\rho} + \frac{\partial \varphi^T}{\partial \mathcal{G}^T}\frac{\partial \mathcal{G}^T}{\partial \rho}\dot{\rho} = -3 \varphi^T_R H(1+w)(1-3w)\kappa\rho +4 \varphi^T_{\mathcal{G}} H(1+w)\Big[ \Lambda (1-3w)  + 2\kappa\rho (1+3w) \Big]\kappa\rho\,.
\end{equation}
The latter can be substituted in Eq.~\eqref{fried} in order to finally arrive at the expression of the reduced modified Friedmann equation, with a cosmological constant $\Lambda$, spatial curvature $k$ and EoS parameter $w$:
\begin{gather}
H^2=\frac{\kappa}{3}\rho - \frac{\Lambda}{3} - \frac{k}{a^2} \nonumber \\
+ \frac{\epsilon}{18} \Bigg\{\frac{6k}{a^2} \kappa \rho  (1+w)  \bigg[16 \varphi^T_{\mathcal{G}\mathcal{G}} (\kappa  \rho -\Lambda) \Big[\Lambda  (1-3w)+2 \kappa  \rho  (1+3w)\Big]+12 \varphi^T_{R \mathcal{G}} \Big[\kappa  \rho  (1+9w)+2\Lambda (1-3 w)\Big]-9 \varphi^T_{R R}
   (1-3w)\bigg]  \nonumber \\
    -  3 \varphi^T +2 (\Lambda -\kappa  \rho ) \Bigg(2 \varphi^T_{\mathcal{G}} \Big[2 \Lambda +\kappa  \rho  (1+3 w)\Big]+\kappa  \rho  (1+w) \bigg[16 \varphi^T_{\mathcal{G}\mathcal{G}} (\kappa  \rho -\Lambda) \Big[\Lambda  (1-3 w)+2
   \kappa  \rho  (1+3 w)\Big]  \nonumber \\
   +12 \varphi^T_{R \mathcal{G}} \Big[ \kappa  \rho  (1+9 w)+2 \Lambda  (1-3 w)\Big]-9 \varphi^T_{R R} (1-3w)\bigg]\Bigg) -3 \varphi^T_{R} \Big[2 \Lambda +\kappa  \rho  (1+3 w)\Big] \Bigg\}. \label{eq1}
\end{gather}
In the equation above we have also replaced $H^2$ by its zero-th order counterpart, {\it i.e.}, $H^2=\frac{1}{3} \kappa \rho -\frac{k}{a^2}-\frac{\Lambda}{3}$.

\section{Bouncing cosmology}
\label{bouncing}

Particularizing the cosmology to a flat spacetime ($k=0$) (in which case the second line of Eq.~\eqref{eq1} vanishes) without a cosmological constant ($\Lambda=0$), Eq.~\eqref{eq1} can be recast into:
\begin{gather}
H^2=\frac{\kappa  \rho }{3} -\frac{\epsilon}{3}\Bigg[ \frac{1}{2}\varphi^T +\frac{2}{3}(1+3w) \kappa ^2 \rho ^2  \varphi^T_{\mathcal{G}} + \frac{32}{3}(1+3w)(1+w) \kappa ^4 \rho ^4  \varphi^T_{\mathcal{G}\mathcal{G}} +\frac{1}{2}(1+3w) \kappa  \rho \varphi^T_{R} \nonumber \\
-3(1-3w)(1+w)\kappa ^2 \rho ^2 \varphi^T_{RR} +4(1+9w)(1+w) \kappa ^3 \rho ^3 \varphi^T_{R\mathcal{G}}  \Bigg], \label{eq2}
\end{gather}
and the scalar curvature invariants given by Eqs.~\eqref{ricciOR} and~\eqref{GBOR} become, respectively:
\begin{eqnarray}
R^T &=&  \left(1-3w \right) \kappa\rho\,,  \label{ricciOR2} \\
\mathcal{G}^T &=& -\frac{4}{3} \left( 1+3w \right)\kappa^2\rho^2\,. \label{GBOR2}
\end{eqnarray}
As we wish to account for the existence of solutions within the framework of $f(R,\mathcal{G})$ gravity, it seems reasonable to look only for solutions where both the curvature scalar $R$ and the invariant $\mathcal{G}$ do not vanish, otherwise our treatment simply reproduces $f(R)$ or $f(\mathcal{G})$ gravity models. Nonetheless, we will comment on how the general solutions at hand reduce to these particular scenarios. Thus, hereafter we will assume $w \neq \pm 1/3$. It is worth mentioning that radiation-like fluids are therefore automatically excluded.

The main purpose of this work is to find an analytical form for the function $f(R,\mathcal{G})$ such that we recover the effective Friedmann equation of a bouncing universe from LQC, which reads
\begin{equation}
\label{qc}
H^2 = \frac{\kappa}{3}\rho \left( 1-\frac{\rho}{\rho_c} \right).
\end{equation}
Since the quantity $\rho_c$ defines the characteristic scale of the problem, and also for dimensionality reasons, it is convenient to define the following dimensionless quantities~\cite{Terrucha:2019jpm}: 
\begin{equation}
\bar{R}^T=\frac{R^T}{\rho_c}\,, \ \  \bar{\mathcal{G}}^T= \frac{\mathcal{G}^T} {\rho_c^2}\,, \ \  \bar{\rho}= \frac{\rho}{\rho_c} \ \  \text{and} \ \  \bar{\varphi}^T=\frac{\varphi^T}{\rho_c}\,.
\label{vardim}
\end{equation}
Then, we can directly equate the right-hand side of the Friedmann Eq.~\eqref{eq2}, to the one we wish to reproduce, that is Eq.~\eqref{qc}. This amounts to a differential equation for $\bar{\varphi}^T(\bar{R}^T,\bar{\mathcal{G}}^T)$ expressed in terms of the dimensionless variables:
\begin{gather}
\frac{1}{2} \bar{\varphi}^T +\frac{2}{3}(1+3w) \kappa ^2 \bar{\rho} ^2  \bar{\varphi}^T_{\bar{\mathcal{G}}} + \frac{32}{3}(1+3w)(1+w) \kappa ^4 \bar{\rho} ^4  \bar{\varphi}^T_{\bar{\mathcal{G}} \bar{\mathcal{G}}} +\frac{1}{2}(1+3w) \kappa  \bar{\rho} \bar{\varphi}^T_{\bar{R}} -3(1-3w)(1+w)\kappa ^2 \bar{\rho} ^2 \bar{\varphi}^T_{\bar{R}\bar{R}} \nonumber \\
 +4(1+9w)(1+w) \kappa ^3 \bar{\rho} ^3 \bar{\varphi}^T_{\bar{R} \bar{\mathcal{G}}} = \frac{\kappa \bar{\rho}^2}{\epsilon}\,,
  \label{eq3res}
\end{gather}
which is a non-linear second-order partial differential equation for a two variable function. We proceed with an empirical approach, by proposing an ansatz for the $f(R,\mathcal{G})$ function (to be more rigorous, for $\bar{\varphi}^T$) as a specific solution of Eq.~\eqref{eq3res}. Afterwards, we investigate the conditions under which the Lagrangian found is capable of reproducing a bouncing universe. We allow the EoS parameter $w$ for the fluid to be arbitrary while in previous works regarding $f(R)$~\cite{Sotiriou:2008ya} and $f(\mathcal{G})$~\cite{Terrucha:2019jpm} gravity only the case $w=1$ was considered.

In the following, we omit the order reduction superscript $T$ for notational convenience. 

\subsection{Solution I}
\label{SpecificI}

The simplest morphology for $\bar{\varphi}$ as a candidate for the solution of the differential equation, Eq.~\eqref{eq3res}, is a power-law form:
\begin{equation}
\label{SolI}
\bar{\varphi} (\bar{R},\bar{\mathcal{G}})= c_1\,|\bar{\mathcal{G}}|^{\alpha}|\bar{R}|^{\beta}\,,
\end{equation}
where $\alpha$, $\beta$ and $c_1$ are real constants. From Eqs.~\eqref{ricciOR2} and~\eqref{GBOR2} we gather that, depending on the value of $w$, the scalars $R$ and $\mathcal{G}$ may become negative. Therefore, in order to avoid the emergence of complex solutions, and without loss of generality, we choose to write the ansatz in terms of the modulus of the scalars, with powers $\{\alpha,\beta\}\in\mathbb{R}$ (this however becomes superfluous for $\lbrace \alpha, \beta \rbrace\in \mathbb{Z}$). Particular cases of this form have been widely considered in the literature to study the stability of cosmological solutions~\cite{delaCruzDombriz:2011wn}, and to describe both inflationary~\cite{DeLaurentis:2015fea} and dark energy~\cite{Alimohammadi:2009js,Bamba:2009uf} scenarios. The particular case $(\alpha, \beta) = (0, 2)$ has already been shown~\cite{Sotiriou:2008ya} to successfully reproduce a cosmological bounce for an $f(R)$ gravity scenario under the same order reduction technique employed in the present manuscript.

Inserting the expression for $\bar{\varphi}$ given in Eq.~\eqref{SolI} directly into the left-hand side of Eq.~\eqref{eq3res}, and making use of the identities in Eqs.~\eqref{ricciOR2}-\eqref{GBOR2}, it follows that:
\begin{equation}
\bar{\rho}^{2\alpha+\beta}\propto \bar{\rho}^2\,.
\label{spec1eq}
\end{equation}
In order for the relation above to be consistent we find that $2\alpha+\beta = 2$ must hold, or equivalently, $\beta=2-2\alpha$. Moreover, the Friedmann equation of a bouncing universe emerging from LQC, that is Eq.~\eqref{qc}, is reproduced from the requirement that the constant $c_1$ reads:
\begin{equation}
\label{C1}
c_1 = \frac{2}{9}\frac{3^{\alpha}}{4^{\alpha}}\frac{1}{(1-\alpha)\epsilon\kappa}\frac{|1-3w|^{2\alpha}|1+3w|^{-\alpha}}{(w+1)(3w-1)}\,.
\end{equation}
Hence, we find a family of solutions of the form,
\begin{equation}
\label{SolIS}
\mathcal{L}_g = R + \frac{C_1}{\rho_c}\,|\mathcal{G}|^{\alpha}|R|^{2-2\alpha}\,,
\end{equation}
with $C_1=\epsilon\, c_1$. The particular cases of $\alpha=1$ or $w=-1$ are excluded as they do not yield a cosmological bounce. When $\alpha=0$, the result found for $f(R)$ gravity (with $w=1$) in Ref.~\cite{Sotiriou:2008ya} is extended to arbitrary values of $w$:
\begin{equation}
\label{SolI2}
\mathcal{L}_g = R+\frac{2}{9 (w+1)(3w-1) \kappa \rho_c}  R^2\,.
\end{equation}
Recall that the values $w=\pm1/3$ were ruled out {\it a priori}. The particular case $w=1/3$ yields $R^T = 0$, from which we require that $\alpha=1$ so that $\varphi \neq 0$. However, this does not feature a cosmological bounce since it leads to $\varphi \propto G$, which is just a topological term. On the other hand, $w=-1/3$ ($\mathcal{G}^T = 0$) is indeed valid provided that $\alpha=0$ and, according to Eq.~\eqref{SolI2}, the correction to the Friedmann equation stems from an $R^2$ term.
 
It is possible to verify that by taking the ansatz $\bar{\varphi}(\bar{R},\bar{\mathcal{G}})=\bar{R} \Psi (\bar{\mathcal{G}})$, and solving Eq.~\eqref{eq3res} for $\Psi$, we obtain a solution where the bounce is induced by Eq.~\eqref{SolIS} with $\alpha=1/2$.

\subsection{Solution II}
\label{SpecificII}

Having attained a solution which encapsulates the model presented in Ref.~\cite{Sotiriou:2008ya}, for an arbitrary $w$, at this point we aim at constructing an ansatz which enclosures the solution found in the framework of $f(\mathcal{G})$ gravity~\cite{Terrucha:2019jpm}. This suggests looking for $\bar{\varphi}$ functions of the following form:
\begin{equation}
\label{SolII}
\bar{\varphi} (\bar{R},\bar{\mathcal{G}})=c_2\,\mathcal{\bar{G}}\ln(|\mathcal{\bar{G}}|^{\beta}|\bar{R}|^{\gamma})\,,
\end{equation}
where $\beta$, $\gamma$ and $c_2$ are real-valued constants. 

Following the same reasoning as in the previous case, we have opted for defining $\bar{\varphi}$ in terms of the modulus of the scalars in order to ensure that the solutions are real-valued and well-defined. This scenario is successful in reproducing the effective Friedmann equation from LQC. We have verified that the term multiplying the logarithm in Eq.~\eqref{SolII} must be exactly $\mathcal{G}$ (for $w\neq -1$) and not some combination of different powers of $R$ and $\mathcal{G}$. This is intricately related to the form of the differential Eq.~\eqref{eq3res} and the factors needed in order to cancel out the dependence on the logarithm in the Friedmann equation. The powers inside the logarithm have no influence whatsoever on the exponent of $\rho$ in the modified Friedmann equation since Eq.~\eqref{SolII} can be manipulated such that they appear merely as multiplicative factors.

Directly inserting the expression for $\bar{\varphi}$, Eq.~\eqref{SolII}, in the differential equation~\eqref{eq3res} yields:
\begin{equation}
\label{Ieq}
c_2 = \frac{3}{2} \frac{1}{\epsilon\kappa}\frac{1-3w}{\beta(-11+24w+27w^2)+\gamma(-1+30w+27w^2)}\,.
\end{equation}
The parameters $\beta$, $\gamma$ and $w$ must be such that the denominator of the last expression is different from zero. The gravitational Lagrangian found is:
\begin{equation}
\label{SolIIS}
\mathcal{L}_g = R+\frac{C_2}{\rho_c} \,\mathcal{G}\ln \frac{|\mathcal{G}|^{\beta}|R|^{\gamma}}{\rho_c^{2\beta+\gamma}}\, ,
\end{equation}
with $C_2 = \epsilon\, c_2$. Solving Eq.~\eqref{eq3res} for the case $\bar{\varphi} (\bar{R},\bar{\mathcal{G}})=\bar{\mathcal{G}}\Psi(\bar{R})$ yields a solution where the bounce is driven by the corrective term in Eq.~\eqref{SolIIS} with $\beta=0$. It is also worth mentioning that the choice $\gamma=0$ corresponds to a generalization of the solution found for $f(\mathcal{G})$ gravity in the case of $w=1$~\cite{Terrucha:2019jpm} to arbitrary values of $w$:
\begin{equation}
\label{perticularID}
\mathcal{L}_g = R-\frac{3}{2 \left(11+9w\right) \kappa \rho_c} \mathcal{G} \ln \frac{|\mathcal{G}|}{\rho_c^2}\,.
\end{equation}
Note that the case $w=1/3$ ($R^T = 0$) is valid for $\gamma = 0$ which leads to $f=f(\mathcal{G})$ governed by Eq.~\eqref{perticularID}. On the other hand, the choice $w=-1/3$ ($\mathcal{G}^T = 0$) is obviously not allowed by the ansatz in Eq.~\eqref{SolII}. As a final remark we briefly mention that, for the particular choice of $w=-1$, it is possible to have a general combination of powers of $R$ and $\mathcal{G}$ multiplying the logarithm and still obtain solutions for the differential equation~\eqref{eq3res}. Thus, exclusively for the case $w=-1$, the gravitational Lagrangian can be extended to:
\begin{equation}
\mathcal{L}_g = R-\frac{6^{\delta}}{4}\frac{1}{\kappa\rho_c(2\beta+\gamma)} \,|\mathcal{G}|^{\delta}|R|^{2-2\delta}\ln \frac{|\mathcal{G}|^{\beta}|R|^{\gamma}}{\rho_c^{2\beta+\gamma}}\,,
\end{equation}
where $\lbrace\beta, \gamma,  \delta\rbrace \in\mathbb{R}$ such that $2\beta+\gamma\neq0$\,.

\section{Conclusions}
\label{conclusions}

In this work we have reconstructed gravitational Lagrangians for $f(R, \mathcal{G})$ gravity that reproduce the Friedmann equation of a bouncing universe, emerging from an effective description of LQC. 

We have formally derived covariant actions for $f(R,\mathcal{G})$ gravity on the basis of an effective field theory approach, which has been implemented by virtue of the order reduction technique. As an artifact of this procedure we were able to find solutions which are perturbatively close to GR while avoiding the, possibly ghost-like, spurious degrees of freedom, generically arising in these theories. Such an approach can lead to much different and richer phenomenological implications than GR such as, in this particular case, to bouncing solutions found in LQC formulations, at early times and near to the Planck scale. 

In this framework, the most general gravitational Lagrangian found can be written as:
\begin{eqnarray}
\label{concEq}
\mathcal{L}_g &=& R+\frac{\mathcal{A}}{\rho_c}\,|\mathcal{G}|^{\alpha}|R|^{2-2\alpha}+\frac{\mathcal{B}}{\rho_c}\,\mathcal{G} \ln \frac{|\mathcal{G}|^{\beta}|R|^{\gamma}}{\rho_c^{2\beta + \gamma}}\,,
\end{eqnarray}
where $\{\mathcal{A},\mathcal{B},\alpha,\beta,\gamma\}\in\mathbb{R}$ are free parameters. The Friedmann equation emerging from the Lagrangian in Eq.~\eqref{concEq} {\it via} the order reduction technique is:
\begin{equation}
H^2 = \frac{\kappa}{3}\rho \left[ 1-\left( \frac{\mathcal{A}}{C_1}+\frac{\mathcal{B}}{C_2} \right)\frac{\rho}{\rho_c} \right].
\label{friedtot}
\end{equation}
In order for the Friedmann equation~\eqref{friedtot} to match the one emerging from LQC, the following constraint has to be satisfied: $\mathcal{A}/C_1+\mathcal{B}/C_2=1$ (hence, only four parameters are indeed free).
Notice that, in order to reproduce the additional term $\propto \rho^2$ in the Friedmann equation, all our solutions satisfy $\epsilon\varphi \sim R^2/\rho_c$ (since $\mathcal{G}\sim R^2$). Thus, as previously mentioned in Sec.~\ref{model}, it is now straightforward to verify that the necessary condition for the validity of the order reduction approach, {\it i.e.} $\epsilon\varphi \ll R$, leads to $R \ll \rho_c$, since $\epsilon\varphi \sim R^2/\rho_c \ll R \, \Rightarrow \, R \ll \rho_c$.

The Lagrangian in Eq.~\eqref{concEq} extends previous works concerning $f(R)$~\cite{Sotiriou:2008ya} and $f(\mathcal{G})$~\cite{Terrucha:2019jpm} gravity, for a fluid with an arbitrary EoS parameter $w$. Let us stress that we have not assumed any specific form for the $f(R,\mathcal{G})$ gravitational action {\it a priori}, but we have found it just as a result of the order reduction technique employed. As a further extension of this work, it would be interesting to investigate if the generalized Friedmann equations obtained in LQC (see Ref.~\cite{Li:2019ipm} and references therein) leading to a bounce, can be obtained in the context of $f(R,\mathcal{G})$ gravity by means of the same order reduction approach.

Finally, it should be noted that since the dynamics of this model deviates from GR at high energies, the primordial power spectrum of both scalar and tensor perturbations is expected to be altered, since it directly depends on the Hubble expansion rate. Such phenomena have also been studied in the context of other theories which present deviations from the standard Friedmann expansion at early times, {\it e.g.} in Randall Sundrum II braneworld model~\cite{Langlois:2000ns,Maartens:1999hf,Barros:2015evi}. Hence, in order to test the robustness of the theory it would be interesting to explore the effects stemming from Eq.~\eqref{friedtot} on the power spectrum of primordial fluctuations and compare them with current observational data.

\acknowledgments
This research was supported by Funda\c{c}\~ao para a Ci\^encia e a Tecnologia (FCT) through the research grant UID/FIS/04434/2019, by the projects PTDC/FIS-OUT/29048/2017, COMPETE2020: POCI-01-0145-FEDER-028987 $\&$ FCT: PTDC/FIS-AST/28987/2017. BJB was supported by the grant PD/BD/128018/2016 from FCT. EMT was supported by the grant BI-MESTRE-IF/00852/2015 from FCT. DV was supported by the grant BI-DOUTOR-IF/00852/2015 from FCT.

\bibliography{bib1}

\begin{thebibliography}{40}%
\makeatletter
\providecommand \@ifxundefined [1]{%
 \@ifx{#1\undefined}
}%
\providecommand \@ifnum [1]{%
 \ifnum #1\expandafter \@firstoftwo
 \else \expandafter \@secondoftwo
 \fi
}%
\providecommand \@ifx [1]{%
 \ifx #1\expandafter \@firstoftwo
 \else \expandafter \@secondoftwo
 \fi
}%
\providecommand \natexlab [1]{#1}%
\providecommand \enquote  [1]{``#1''}%
\providecommand \bibnamefont  [1]{#1}%
\providecommand \bibfnamefont [1]{#1}%
\providecommand \citenamefont [1]{#1}%
\providecommand \href@noop [0]{\@secondoftwo}%
\providecommand \href [0]{\begingroup \@sanitize@url \@href}%
\providecommand \@href[1]{\@@startlink{#1}\@@href}%
\providecommand \@@href[1]{\endgroup#1\@@endlink}%
\providecommand \@sanitize@url [0]{\catcode `\\12\catcode `\$12\catcode
  `\&12\catcode `\#12\catcode `\^12\catcode `\_12\catcode `\%12\relax}%
\providecommand \@@startlink[1]{}%
\providecommand \@@endlink[0]{}%
\providecommand \url  [0]{\begingroup\@sanitize@url \@url }%
\providecommand \@url [1]{\endgroup\@href {#1}{\urlprefix }}%
\providecommand \urlprefix  [0]{URL }%
\providecommand \Eprint [0]{\href }%
\providecommand \doibase [0]{http://dx.doi.org/}%
\providecommand \selectlanguage [0]{\@gobble}%
\providecommand \bibinfo  [0]{\@secondoftwo}%
\providecommand \bibfield  [0]{\@secondoftwo}%
\providecommand \translation [1]{[#1]}%
\providecommand \BibitemOpen [0]{}%
\providecommand \bibitemStop [0]{}%
\providecommand \bibitemNoStop [0]{.\EOS\space}%
\providecommand \EOS [0]{\spacefactor3000\relax}%
\providecommand \BibitemShut  [1]{\csname bibitem#1\endcsname}%
\let\auto@bib@innerbib\@empty
\bibitem [{\citenamefont {Bojowald}(2005)}]{Bojowald:2006da}%
  \BibitemOpen
  \bibfield  {author} {\bibinfo {author} {\bibfnamefont {M.}~\bibnamefont
  {Bojowald}},\ }\href {\doibase 10.12942/lrr-2005-11} {\bibfield  {journal}
  {\bibinfo  {journal} {Living Rev. Rel.}\ }\textbf {\bibinfo {volume} {8}},\
  \bibinfo {pages} {11} (\bibinfo {year} {2005})},\ \Eprint
  {http://arxiv.org/abs/gr-qc/0601085} {arXiv:gr-qc/0601085 [gr-qc]}
  \BibitemShut {NoStop}%
\bibitem [{\citenamefont {Ashtekar}\ and\ \citenamefont
  {Singh}(2011)}]{Ashtekar:2011ni}%
  \BibitemOpen
  \bibfield  {author} {\bibinfo {author} {\bibfnamefont {A.}~\bibnamefont
  {Ashtekar}}\ and\ \bibinfo {author} {\bibfnamefont {P.}~\bibnamefont
  {Singh}},\ }\href {\doibase 10.1088/0264-9381/28/21/213001} {\bibfield
  {journal} {\bibinfo  {journal} {Class. Quant. Grav.}\ }\textbf {\bibinfo
  {volume} {28}},\ \bibinfo {pages} {213001} (\bibinfo {year} {2011})},\
  \Eprint {http://arxiv.org/abs/1108.0893} {arXiv:1108.0893 [gr-qc]}
  \BibitemShut {NoStop}%
\bibitem [{\citenamefont {Tsujikawa}\ \emph {et~al.}(2004)\citenamefont
  {Tsujikawa}, \citenamefont {Singh},\ and\ \citenamefont
  {Maartens}}]{Tsujikawa:2003vr}%
  \BibitemOpen
  \bibfield  {author} {\bibinfo {author} {\bibfnamefont {S.}~\bibnamefont
  {Tsujikawa}}, \bibinfo {author} {\bibfnamefont {P.}~\bibnamefont {Singh}}, \
  and\ \bibinfo {author} {\bibfnamefont {R.}~\bibnamefont {Maartens}},\ }\href
  {\doibase 10.1088/0264-9381/21/24/006} {\bibfield  {journal} {\bibinfo
  {journal} {Class. Quant. Grav.}\ }\textbf {\bibinfo {volume} {21}},\ \bibinfo
  {pages} {5767} (\bibinfo {year} {2004})},\ \Eprint
  {http://arxiv.org/abs/astro-ph/0311015} {arXiv:astro-ph/0311015} \BibitemShut
  {NoStop}%
\bibitem [{\citenamefont {Bojowald}\ \emph
  {et~al.}(2004{\natexlab{a}})\citenamefont {Bojowald}, \citenamefont {Lidsey},
  \citenamefont {Mulryne}, \citenamefont {Singh},\ and\ \citenamefont
  {Tavakol}}]{Bojowald:2004xq}%
  \BibitemOpen
  \bibfield  {author} {\bibinfo {author} {\bibfnamefont {M.}~\bibnamefont
  {Bojowald}}, \bibinfo {author} {\bibfnamefont {J.~E.}\ \bibnamefont
  {Lidsey}}, \bibinfo {author} {\bibfnamefont {D.~J.}\ \bibnamefont {Mulryne}},
  \bibinfo {author} {\bibfnamefont {P.}~\bibnamefont {Singh}}, \ and\ \bibinfo
  {author} {\bibfnamefont {R.}~\bibnamefont {Tavakol}},\ }\href {\doibase
  10.1103/PhysRevD.70.043530} {\bibfield  {journal} {\bibinfo  {journal} {Phys.
  Rev. D}\ }\textbf {\bibinfo {volume} {70}},\ \bibinfo {pages} {043530}
  (\bibinfo {year} {2004}{\natexlab{a}})},\ \Eprint
  {http://arxiv.org/abs/gr-qc/0403106} {arXiv:gr-qc/0403106} \BibitemShut
  {NoStop}%
\bibitem [{\citenamefont {Singh}\ and\ \citenamefont
  {Toporensky}(2004)}]{Singh:2003au}%
  \BibitemOpen
  \bibfield  {author} {\bibinfo {author} {\bibfnamefont {P.}~\bibnamefont
  {Singh}}\ and\ \bibinfo {author} {\bibfnamefont {A.}~\bibnamefont
  {Toporensky}},\ }\href {\doibase 10.1103/PhysRevD.69.104008} {\bibfield
  {journal} {\bibinfo  {journal} {Phys. Rev. D}\ }\textbf {\bibinfo {volume}
  {69}},\ \bibinfo {pages} {104008} (\bibinfo {year} {2004})},\ \Eprint
  {http://arxiv.org/abs/gr-qc/0312110} {arXiv:gr-qc/0312110} \BibitemShut
  {NoStop}%
\bibitem [{\citenamefont {Bojowald}(2001)}]{Bojowald:2001xe}%
  \BibitemOpen
  \bibfield  {author} {\bibinfo {author} {\bibfnamefont {M.}~\bibnamefont
  {Bojowald}},\ }\href {\doibase 10.1103/PhysRevLett.86.5227} {\bibfield
  {journal} {\bibinfo  {journal} {Phys. Rev. Lett.}\ }\textbf {\bibinfo
  {volume} {86}},\ \bibinfo {pages} {5227} (\bibinfo {year} {2001})},\ \Eprint
  {http://arxiv.org/abs/gr-qc/0102069} {arXiv:gr-qc/0102069 [gr-qc]}
  \BibitemShut {NoStop}%
\bibitem [{\citenamefont {Bojowald}\ \emph
  {et~al.}(2004{\natexlab{b}})\citenamefont {Bojowald}, \citenamefont
  {Maartens},\ and\ \citenamefont {Singh}}]{Bojowald:2004kt}%
  \BibitemOpen
  \bibfield  {author} {\bibinfo {author} {\bibfnamefont {M.}~\bibnamefont
  {Bojowald}}, \bibinfo {author} {\bibfnamefont {R.}~\bibnamefont {Maartens}},
  \ and\ \bibinfo {author} {\bibfnamefont {P.}~\bibnamefont {Singh}},\ }\href
  {\doibase 10.1103/PhysRevD.70.083517} {\bibfield  {journal} {\bibinfo
  {journal} {Phys. Rev. D}\ }\textbf {\bibinfo {volume} {70}},\ \bibinfo
  {pages} {083517} (\bibinfo {year} {2004}{\natexlab{b}})},\ \Eprint
  {http://arxiv.org/abs/hep-th/0407115} {arXiv:hep-th/0407115} \BibitemShut
  {NoStop}%
\bibitem [{\citenamefont {Ashtekar}\ \emph {et~al.}(2006)\citenamefont
  {Ashtekar}, \citenamefont {Pawlowski},\ and\ \citenamefont
  {Singh}}]{Ashtekar:2006wn}%
  \BibitemOpen
  \bibfield  {author} {\bibinfo {author} {\bibfnamefont {A.}~\bibnamefont
  {Ashtekar}}, \bibinfo {author} {\bibfnamefont {T.}~\bibnamefont {Pawlowski}},
  \ and\ \bibinfo {author} {\bibfnamefont {P.}~\bibnamefont {Singh}},\ }\href
  {\doibase 10.1103/PhysRevD.74.084003} {\bibfield  {journal} {\bibinfo
  {journal} {Phys. Rev.}\ }\textbf {\bibinfo {volume} {D74}},\ \bibinfo {pages}
  {084003} (\bibinfo {year} {2006})},\ \Eprint
  {http://arxiv.org/abs/gr-qc/0607039} {arXiv:gr-qc/0607039 [gr-qc]}
  \BibitemShut {NoStop}%
\bibitem [{\citenamefont {Brandenberger}(2012)}]{Brandenberger:2012zb}%
  \BibitemOpen
  \bibfield  {author} {\bibinfo {author} {\bibfnamefont {R.~H.}\ \bibnamefont
  {Brandenberger}},\ }\href@noop {} {\  (\bibinfo {year} {2012})},\ \Eprint
  {http://arxiv.org/abs/1206.4196} {arXiv:1206.4196 [astro-ph.CO]} \BibitemShut
  {NoStop}%
\bibitem [{\citenamefont {Odintsov}\ and\ \citenamefont
  {Oikonomou}(2014)}]{Odintsov:2014gea}%
  \BibitemOpen
  \bibfield  {author} {\bibinfo {author} {\bibfnamefont {S.}~\bibnamefont
  {Odintsov}}\ and\ \bibinfo {author} {\bibfnamefont {V.}~\bibnamefont
  {Oikonomou}},\ }\href {\doibase 10.1103/PhysRevD.90.124083} {\bibfield
  {journal} {\bibinfo  {journal} {Phys. Rev. D}\ }\textbf {\bibinfo {volume}
  {90}},\ \bibinfo {pages} {124083} (\bibinfo {year} {2014})},\ \Eprint
  {http://arxiv.org/abs/1410.8183} {arXiv:1410.8183 [gr-qc]} \BibitemShut
  {NoStop}%
\bibitem [{\citenamefont {Odintsov}\ and\ \citenamefont
  {Oikonomou}(2017)}]{Odintsov:2015ynk}%
  \BibitemOpen
  \bibfield  {author} {\bibinfo {author} {\bibfnamefont {S.}~\bibnamefont
  {Odintsov}}\ and\ \bibinfo {author} {\bibfnamefont {V.}~\bibnamefont
  {Oikonomou}},\ }\href {\doibase 10.1142/S0218271817500857} {\bibfield
  {journal} {\bibinfo  {journal} {Int. J. Mod. Phys. D}\ }\textbf {\bibinfo
  {volume} {26}},\ \bibinfo {pages} {1750085} (\bibinfo {year} {2017})},\
  \Eprint {http://arxiv.org/abs/1512.04787} {arXiv:1512.04787 [gr-qc]}
  \BibitemShut {NoStop}%
\bibitem [{\citenamefont {Odintsov}\ and\ \citenamefont
  {Oikonomou}(2016)}]{Odintsov:2016tar}%
  \BibitemOpen
  \bibfield  {author} {\bibinfo {author} {\bibfnamefont {S.}~\bibnamefont
  {Odintsov}}\ and\ \bibinfo {author} {\bibfnamefont {V.}~\bibnamefont
  {Oikonomou}},\ }\href {\doibase 10.1103/PhysRevD.94.064022} {\bibfield
  {journal} {\bibinfo  {journal} {Phys. Rev. D}\ }\textbf {\bibinfo {volume}
  {94}},\ \bibinfo {pages} {064022} (\bibinfo {year} {2016})},\ \Eprint
  {http://arxiv.org/abs/1606.03689} {arXiv:1606.03689 [gr-qc]} \BibitemShut
  {NoStop}%
\bibitem [{\citenamefont {Taveras}(2008)}]{Taveras:2008ke}%
  \BibitemOpen
  \bibfield  {author} {\bibinfo {author} {\bibfnamefont {V.}~\bibnamefont
  {Taveras}},\ }\href {\doibase 10.1103/PhysRevD.78.064072} {\bibfield
  {journal} {\bibinfo  {journal} {Phys. Rev.}\ }\textbf {\bibinfo {volume}
  {D78}},\ \bibinfo {pages} {064072} (\bibinfo {year} {2008})},\ \Eprint
  {http://arxiv.org/abs/0807.3325} {arXiv:0807.3325 [gr-qc]} \BibitemShut
  {NoStop}%
\bibitem [{\citenamefont {Banerjee}\ \emph {et~al.}(2012)\citenamefont
  {Banerjee}, \citenamefont {Calcagni},\ and\ \citenamefont
  {Martin-Benito}}]{Banerjee:2011qu}%
  \BibitemOpen
  \bibfield  {author} {\bibinfo {author} {\bibfnamefont {K.}~\bibnamefont
  {Banerjee}}, \bibinfo {author} {\bibfnamefont {G.}~\bibnamefont {Calcagni}},
  \ and\ \bibinfo {author} {\bibfnamefont {M.}~\bibnamefont {Martin-Benito}},\
  }\href {\doibase 10.3842/SIGMA.2012.016} {\bibfield  {journal} {\bibinfo
  {journal} {SIGMA}\ }\textbf {\bibinfo {volume} {8}},\ \bibinfo {pages} {016}
  (\bibinfo {year} {2012})},\ \Eprint {http://arxiv.org/abs/1109.6801}
  {arXiv:1109.6801 [gr-qc]} \BibitemShut {NoStop}%
\bibitem [{\citenamefont {Capozziello}\ and\ \citenamefont
  {Francaviglia}(2008)}]{Capozziello:2007ec}%
  \BibitemOpen
  \bibfield  {author} {\bibinfo {author} {\bibfnamefont {S.}~\bibnamefont
  {Capozziello}}\ and\ \bibinfo {author} {\bibfnamefont {M.}~\bibnamefont
  {Francaviglia}},\ }\href {\doibase 10.1007/s10714-007-0551-y} {\bibfield
  {journal} {\bibinfo  {journal} {Gen. Rel. Grav.}\ }\textbf {\bibinfo {volume}
  {40}},\ \bibinfo {pages} {357} (\bibinfo {year} {2008})},\ \Eprint
  {http://arxiv.org/abs/0706.1146} {arXiv:0706.1146 [astro-ph]} \BibitemShut
  {NoStop}%
\bibitem [{\citenamefont {Nojiri}\ \emph {et~al.}(2017)\citenamefont {Nojiri},
  \citenamefont {Odintsov},\ and\ \citenamefont {Oikonomou}}]{Nojiri:2017ncd}%
  \BibitemOpen
  \bibfield  {author} {\bibinfo {author} {\bibfnamefont {S.}~\bibnamefont
  {Nojiri}}, \bibinfo {author} {\bibfnamefont {S.}~\bibnamefont {Odintsov}}, \
  and\ \bibinfo {author} {\bibfnamefont {V.}~\bibnamefont {Oikonomou}},\ }\href
  {\doibase 10.1016/j.physrep.2017.06.001} {\bibfield  {journal} {\bibinfo
  {journal} {Phys. Rept.}\ }\textbf {\bibinfo {volume} {692}},\ \bibinfo
  {pages} {1} (\bibinfo {year} {2017})},\ \Eprint
  {http://arxiv.org/abs/1705.11098} {arXiv:1705.11098 [gr-qc]} \BibitemShut
  {NoStop}%
\bibitem [{\citenamefont {De~Felice}\ and\ \citenamefont
  {Suyama}(2009)}]{DeFelice:2009ak}%
  \BibitemOpen
  \bibfield  {author} {\bibinfo {author} {\bibfnamefont {A.}~\bibnamefont
  {De~Felice}}\ and\ \bibinfo {author} {\bibfnamefont {T.}~\bibnamefont
  {Suyama}},\ }\href {\doibase 10.1088/1475-7516/2009/06/034} {\bibfield
  {journal} {\bibinfo  {journal} {JCAP}\ }\textbf {\bibinfo {volume} {0906}},\
  \bibinfo {pages} {034} (\bibinfo {year} {2009})},\ \Eprint
  {http://arxiv.org/abs/0904.2092} {arXiv:0904.2092 [astro-ph.CO]} \BibitemShut
  {NoStop}%
\bibitem [{\citenamefont {Odintsov}\ \emph {et~al.}(2019)\citenamefont
  {Odintsov}, \citenamefont {Oikonomou},\ and\ \citenamefont
  {Banerjee}}]{Odintsov:2018nch}%
  \BibitemOpen
  \bibfield  {author} {\bibinfo {author} {\bibfnamefont {S.~D.}\ \bibnamefont
  {Odintsov}}, \bibinfo {author} {\bibfnamefont {V.~K.}\ \bibnamefont
  {Oikonomou}}, \ and\ \bibinfo {author} {\bibfnamefont {S.}~\bibnamefont
  {Banerjee}},\ }\href {\doibase 10.1016/j.nuclphysb.2018.07.013} {\bibfield
  {journal} {\bibinfo  {journal} {Nucl. Phys.}\ }\textbf {\bibinfo {volume}
  {B938}},\ \bibinfo {pages} {935} (\bibinfo {year} {2019})},\ \Eprint
  {http://arxiv.org/abs/1807.00335} {arXiv:1807.00335 [gr-qc]} \BibitemShut
  {NoStop}%
\bibitem [{\citenamefont {Odintsov}\ \emph {et~al.}(2020)\citenamefont
  {Odintsov}, \citenamefont {Oikonomou},\ and\ \citenamefont
  {Fronimos}}]{Odintsov:2020sqy}%
  \BibitemOpen
  \bibfield  {author} {\bibinfo {author} {\bibfnamefont {S.}~\bibnamefont
  {Odintsov}}, \bibinfo {author} {\bibfnamefont {V.}~\bibnamefont {Oikonomou}},
  \ and\ \bibinfo {author} {\bibfnamefont {F.}~\bibnamefont {Fronimos}},\
  }\href@noop {} {\  (\bibinfo {year} {2020})},\ \Eprint
  {http://arxiv.org/abs/2003.13724} {arXiv:2003.13724 [gr-qc]} \BibitemShut
  {NoStop}%
\bibitem [{\citenamefont {Alimohammadi}\ and\ \citenamefont
  {Ghalee}(2009{\natexlab{a}})}]{Alimohammadi:2008fq}%
  \BibitemOpen
  \bibfield  {author} {\bibinfo {author} {\bibfnamefont {M.}~\bibnamefont
  {Alimohammadi}}\ and\ \bibinfo {author} {\bibfnamefont {A.}~\bibnamefont
  {Ghalee}},\ }\href {\doibase 10.1103/PhysRevD.79.063006} {\bibfield
  {journal} {\bibinfo  {journal} {Phys. Rev.}\ }\textbf {\bibinfo {volume}
  {D79}},\ \bibinfo {pages} {063006} (\bibinfo {year} {2009}{\natexlab{a}})},\
  \Eprint {http://arxiv.org/abs/0811.1286} {arXiv:0811.1286 [gr-qc]}
  \BibitemShut {NoStop}%
\bibitem [{\citenamefont {Elizalde}\ \emph {et~al.}(2010)\citenamefont
  {Elizalde}, \citenamefont {Myrzakulov}, \citenamefont {Obukhov},\ and\
  \citenamefont {Saez-Gomez}}]{Elizalde:2010jx}%
  \BibitemOpen
  \bibfield  {author} {\bibinfo {author} {\bibfnamefont {E.}~\bibnamefont
  {Elizalde}}, \bibinfo {author} {\bibfnamefont {R.}~\bibnamefont
  {Myrzakulov}}, \bibinfo {author} {\bibfnamefont {V.~V.}\ \bibnamefont
  {Obukhov}}, \ and\ \bibinfo {author} {\bibfnamefont {D.}~\bibnamefont
  {Saez-Gomez}},\ }\href {\doibase 10.1088/0264-9381/27/9/095007} {\bibfield
  {journal} {\bibinfo  {journal} {Class. Quant. Grav.}\ }\textbf {\bibinfo
  {volume} {27}},\ \bibinfo {pages} {095007} (\bibinfo {year} {2010})},\
  \Eprint {http://arxiv.org/abs/1001.3636} {arXiv:1001.3636 [gr-qc]}
  \BibitemShut {NoStop}%
\bibitem [{\citenamefont {Bamba}\ \emph {et~al.}(2010)\citenamefont {Bamba},
  \citenamefont {Odintsov}, \citenamefont {Sebastiani},\ and\ \citenamefont
  {Zerbini}}]{Bamba:2009uf}%
  \BibitemOpen
  \bibfield  {author} {\bibinfo {author} {\bibfnamefont {K.}~\bibnamefont
  {Bamba}}, \bibinfo {author} {\bibfnamefont {S.~D.}\ \bibnamefont {Odintsov}},
  \bibinfo {author} {\bibfnamefont {L.}~\bibnamefont {Sebastiani}}, \ and\
  \bibinfo {author} {\bibfnamefont {S.}~\bibnamefont {Zerbini}},\ }\href
  {\doibase 10.1140/epjc/s10052-010-1292-8} {\bibfield  {journal} {\bibinfo
  {journal} {Eur. Phys. J.}\ }\textbf {\bibinfo {volume} {C67}},\ \bibinfo
  {pages} {295} (\bibinfo {year} {2010})},\ \Eprint
  {http://arxiv.org/abs/0911.4390} {arXiv:0911.4390 [hep-th]} \BibitemShut
  {NoStop}%
\bibitem [{\citenamefont {De~Felice}\ \emph {et~al.}(2010)\citenamefont
  {De~Felice}, \citenamefont {Gerard},\ and\ \citenamefont
  {Suyama}}]{DeFelice:2010sh}%
  \BibitemOpen
  \bibfield  {author} {\bibinfo {author} {\bibfnamefont {A.}~\bibnamefont
  {De~Felice}}, \bibinfo {author} {\bibfnamefont {J.-M.}\ \bibnamefont
  {Gerard}}, \ and\ \bibinfo {author} {\bibfnamefont {T.}~\bibnamefont
  {Suyama}},\ }\href {\doibase 10.1103/PhysRevD.82.063526} {\bibfield
  {journal} {\bibinfo  {journal} {Phys. Rev.}\ }\textbf {\bibinfo {volume}
  {D82}},\ \bibinfo {pages} {063526} (\bibinfo {year} {2010})},\ \Eprint
  {http://arxiv.org/abs/1005.1958} {arXiv:1005.1958 [astro-ph.CO]} \BibitemShut
  {NoStop}%
\bibitem [{\citenamefont {De~Felice}\ and\ \citenamefont
  {Tanaka}(2010)}]{DeFelice:2010hg}%
  \BibitemOpen
  \bibfield  {author} {\bibinfo {author} {\bibfnamefont {A.}~\bibnamefont
  {De~Felice}}\ and\ \bibinfo {author} {\bibfnamefont {T.}~\bibnamefont
  {Tanaka}},\ }\href {\doibase 10.1143/PTP.124.503} {\bibfield  {journal}
  {\bibinfo  {journal} {Prog. Theor. Phys.}\ }\textbf {\bibinfo {volume}
  {124}},\ \bibinfo {pages} {503} (\bibinfo {year} {2010})},\ \Eprint
  {http://arxiv.org/abs/1006.4399} {arXiv:1006.4399 [astro-ph.CO]} \BibitemShut
  {NoStop}%
\bibitem [{\citenamefont {De~Laurentis}\ \emph {et~al.}(2015)\citenamefont
  {De~Laurentis}, \citenamefont {Paolella},\ and\ \citenamefont
  {Capozziello}}]{DeLaurentis:2015fea}%
  \BibitemOpen
  \bibfield  {author} {\bibinfo {author} {\bibfnamefont {M.}~\bibnamefont
  {De~Laurentis}}, \bibinfo {author} {\bibfnamefont {M.}~\bibnamefont
  {Paolella}}, \ and\ \bibinfo {author} {\bibfnamefont {S.}~\bibnamefont
  {Capozziello}},\ }\href {\doibase 10.1103/PhysRevD.91.083531} {\bibfield
  {journal} {\bibinfo  {journal} {Phys. Rev.}\ }\textbf {\bibinfo {volume}
  {D91}},\ \bibinfo {pages} {083531} (\bibinfo {year} {2015})},\ \Eprint
  {http://arxiv.org/abs/1503.04659} {arXiv:1503.04659 [gr-qc]} \BibitemShut
  {NoStop}%
\bibitem [{\citenamefont {Oikonomou}(2015)}]{Oikonomou:2015qha}%
  \BibitemOpen
  \bibfield  {author} {\bibinfo {author} {\bibfnamefont {V.}~\bibnamefont
  {Oikonomou}},\ }\href {\doibase 10.1103/PhysRevD.92.124027} {\bibfield
  {journal} {\bibinfo  {journal} {Phys. Rev. D}\ }\textbf {\bibinfo {volume}
  {92}},\ \bibinfo {pages} {124027} (\bibinfo {year} {2015})},\ \Eprint
  {http://arxiv.org/abs/1509.05827} {arXiv:1509.05827 [gr-qc]} \BibitemShut
  {NoStop}%
\bibitem [{\citenamefont {Li}\ \emph {et~al.}(2007)\citenamefont {Li},
  \citenamefont {Barrow},\ and\ \citenamefont {Mota}}]{Li:2007jm}%
  \BibitemOpen
  \bibfield  {author} {\bibinfo {author} {\bibfnamefont {B.}~\bibnamefont
  {Li}}, \bibinfo {author} {\bibfnamefont {J.~D.}\ \bibnamefont {Barrow}}, \
  and\ \bibinfo {author} {\bibfnamefont {D.~F.}\ \bibnamefont {Mota}},\ }\href
  {\doibase 10.1103/PhysRevD.76.044027} {\bibfield  {journal} {\bibinfo
  {journal} {Phys. Rev. D}\ }\textbf {\bibinfo {volume} {76}},\ \bibinfo
  {pages} {044027} (\bibinfo {year} {2007})},\ \Eprint
  {http://arxiv.org/abs/0705.3795} {arXiv:0705.3795 [gr-qc]} \BibitemShut
  {NoStop}%
\bibitem [{\citenamefont {Bel}\ and\ \citenamefont {Zia}(1985)}]{Bel:1985zz}%
  \BibitemOpen
  \bibfield  {author} {\bibinfo {author} {\bibfnamefont {L.}~\bibnamefont
  {Bel}}\ and\ \bibinfo {author} {\bibfnamefont {H.~S.}\ \bibnamefont {Zia}},\
  }\href {\doibase 10.1103/PhysRevD.32.3128} {\bibfield  {journal} {\bibinfo
  {journal} {Phys. Rev.}\ }\textbf {\bibinfo {volume} {D32}},\ \bibinfo {pages}
  {3128} (\bibinfo {year} {1985})}\BibitemShut {NoStop}%
\bibitem [{\citenamefont {Simon}(1990)}]{Simon:1990ic}%
  \BibitemOpen
  \bibfield  {author} {\bibinfo {author} {\bibfnamefont {J.~Z.}\ \bibnamefont
  {Simon}},\ }\href {\doibase 10.1103/PhysRevD.41.3720} {\bibfield  {journal}
  {\bibinfo  {journal} {Phys. Rev.}\ }\textbf {\bibinfo {volume} {D41}},\
  \bibinfo {pages} {3720} (\bibinfo {year} {1990})}\BibitemShut {NoStop}%
\bibitem [{\citenamefont {Simon}(1992)}]{Simon:1991bm}%
  \BibitemOpen
  \bibfield  {author} {\bibinfo {author} {\bibfnamefont {J.~Z.}\ \bibnamefont
  {Simon}},\ }\href {\doibase 10.1103/PhysRevD.45.1953} {\bibfield  {journal}
  {\bibinfo  {journal} {Phys. Rev.}\ }\textbf {\bibinfo {volume} {D45}},\
  \bibinfo {pages} {1953} (\bibinfo {year} {1992})}\BibitemShut {NoStop}%
\bibitem [{\citenamefont {Sotiriou}(2009)}]{Sotiriou:2008ya}%
  \BibitemOpen
  \bibfield  {author} {\bibinfo {author} {\bibfnamefont {T.~P.}\ \bibnamefont
  {Sotiriou}},\ }\href {\doibase 10.1103/PhysRevD.79.044035} {\bibfield
  {journal} {\bibinfo  {journal} {Phys. Rev.}\ }\textbf {\bibinfo {volume}
  {D79}},\ \bibinfo {pages} {044035} (\bibinfo {year} {2009})},\ \Eprint
  {http://arxiv.org/abs/0811.1799} {arXiv:0811.1799 [gr-qc]} \BibitemShut
  {NoStop}%
\bibitem [{\citenamefont {Terrucha}\ \emph {et~al.}(2019)\citenamefont
  {Terrucha}, \citenamefont {Vernieri},\ and\ \citenamefont
  {Lemos}}]{Terrucha:2019jpm}%
  \BibitemOpen
  \bibfield  {author} {\bibinfo {author} {\bibfnamefont {I.}~\bibnamefont
  {Terrucha}}, \bibinfo {author} {\bibfnamefont {D.}~\bibnamefont {Vernieri}},
  \ and\ \bibinfo {author} {\bibfnamefont {J.~P.~S.}\ \bibnamefont {Lemos}},\
  }\href {\doibase 10.1016/j.aop.2019.02.010} {\bibfield  {journal} {\bibinfo
  {journal} {Annals Phys.}\ }\textbf {\bibinfo {volume} {404}},\ \bibinfo
  {pages} {39} (\bibinfo {year} {2019})},\ \Eprint
  {http://arxiv.org/abs/1904.00260} {arXiv:1904.00260 [gr-qc]} \BibitemShut
  {NoStop}%
\bibitem [{\citenamefont {Olmo}\ and\ \citenamefont
  {Singh}(2009)}]{Olmo:2008nf}%
  \BibitemOpen
  \bibfield  {author} {\bibinfo {author} {\bibfnamefont {G.~J.}\ \bibnamefont
  {Olmo}}\ and\ \bibinfo {author} {\bibfnamefont {P.}~\bibnamefont {Singh}},\
  }\href {\doibase 10.1088/1475-7516/2009/01/030} {\bibfield  {journal}
  {\bibinfo  {journal} {JCAP}\ }\textbf {\bibinfo {volume} {0901}},\ \bibinfo
  {pages} {030} (\bibinfo {year} {2009})},\ \Eprint
  {http://arxiv.org/abs/0806.2783} {arXiv:0806.2783 [gr-qc]} \BibitemShut
  {NoStop}%
\bibitem [{\citenamefont {Langlois}\ \emph {et~al.}(2017)\citenamefont
  {Langlois}, \citenamefont {Liu}, \citenamefont {Noui},\ and\ \citenamefont
  {Wilson-Ewing}}]{Liu:2017puc}%
  \BibitemOpen
  \bibfield  {author} {\bibinfo {author} {\bibfnamefont {D.}~\bibnamefont
  {Langlois}}, \bibinfo {author} {\bibfnamefont {H.}~\bibnamefont {Liu}},
  \bibinfo {author} {\bibfnamefont {K.}~\bibnamefont {Noui}}, \ and\ \bibinfo
  {author} {\bibfnamefont {E.}~\bibnamefont {Wilson-Ewing}},\ }\href {\doibase
  10.1088/1361-6382/aa8f2f} {\bibfield  {journal} {\bibinfo  {journal} {Class.
  Quant. Grav.}\ }\textbf {\bibinfo {volume} {34}},\ \bibinfo {pages} {225004}
  (\bibinfo {year} {2017})},\ \Eprint {http://arxiv.org/abs/1703.10812}
  {arXiv:1703.10812 [gr-qc]} \BibitemShut {NoStop}%
\bibitem [{\citenamefont {de~la Cruz-Dombriz}\ and\ \citenamefont
  {Saez-Gomez}(2012)}]{delaCruzDombriz:2011wn}%
  \BibitemOpen
  \bibfield  {author} {\bibinfo {author} {\bibfnamefont {A.}~\bibnamefont
  {de~la Cruz-Dombriz}}\ and\ \bibinfo {author} {\bibfnamefont
  {D.}~\bibnamefont {Saez-Gomez}},\ }\href {\doibase
  10.1088/0264-9381/29/24/245014} {\bibfield  {journal} {\bibinfo  {journal}
  {Class. Quant. Grav.}\ }\textbf {\bibinfo {volume} {29}},\ \bibinfo {pages}
  {245014} (\bibinfo {year} {2012})},\ \Eprint {http://arxiv.org/abs/1112.4481}
  {arXiv:1112.4481 [gr-qc]} \BibitemShut {NoStop}%
\bibitem [{\citenamefont {Alimohammadi}\ and\ \citenamefont
  {Ghalee}(2009{\natexlab{b}})}]{Alimohammadi:2009js}%
  \BibitemOpen
  \bibfield  {author} {\bibinfo {author} {\bibfnamefont {M.}~\bibnamefont
  {Alimohammadi}}\ and\ \bibinfo {author} {\bibfnamefont {A.}~\bibnamefont
  {Ghalee}},\ }\href {\doibase 10.1103/PhysRevD.80.043006} {\bibfield
  {journal} {\bibinfo  {journal} {Phys. Rev.}\ }\textbf {\bibinfo {volume}
  {D80}},\ \bibinfo {pages} {043006} (\bibinfo {year} {2009}{\natexlab{b}})},\
  \Eprint {http://arxiv.org/abs/0908.1150} {arXiv:0908.1150 [gr-qc]}
  \BibitemShut {NoStop}%
\bibitem [{\citenamefont {Li}\ \emph {et~al.}(2019)\citenamefont {Li},
  \citenamefont {Singh},\ and\ \citenamefont {Wang}}]{Li:2019ipm}%
  \BibitemOpen
  \bibfield  {author} {\bibinfo {author} {\bibfnamefont {B.-F.}\ \bibnamefont
  {Li}}, \bibinfo {author} {\bibfnamefont {P.}~\bibnamefont {Singh}}, \ and\
  \bibinfo {author} {\bibfnamefont {A.}~\bibnamefont {Wang}},\ }\href {\doibase
  10.1103/PhysRevD.100.063513} {\bibfield  {journal} {\bibinfo  {journal}
  {Phys. Rev.}\ }\textbf {\bibinfo {volume} {D100}},\ \bibinfo {pages} {063513}
  (\bibinfo {year} {2019})},\ \Eprint {http://arxiv.org/abs/1906.01001}
  {arXiv:1906.01001 [gr-qc]} \BibitemShut {NoStop}%
\bibitem [{\citenamefont {Langlois}\ \emph {et~al.}(2000)\citenamefont
  {Langlois}, \citenamefont {Maartens},\ and\ \citenamefont
  {Wands}}]{Langlois:2000ns}%
  \BibitemOpen
  \bibfield  {author} {\bibinfo {author} {\bibfnamefont {D.}~\bibnamefont
  {Langlois}}, \bibinfo {author} {\bibfnamefont {R.}~\bibnamefont {Maartens}},
  \ and\ \bibinfo {author} {\bibfnamefont {D.}~\bibnamefont {Wands}},\ }\href
  {\doibase 10.1016/S0370-2693(00)00957-6} {\bibfield  {journal} {\bibinfo
  {journal} {Phys. Lett. B}\ }\textbf {\bibinfo {volume} {489}},\ \bibinfo
  {pages} {259} (\bibinfo {year} {2000})},\ \Eprint
  {http://arxiv.org/abs/hep-th/0006007} {arXiv:hep-th/0006007} \BibitemShut
  {NoStop}%
\bibitem [{\citenamefont {Maartens}\ \emph {et~al.}(2000)\citenamefont
  {Maartens}, \citenamefont {Wands}, \citenamefont {Bassett},\ and\
  \citenamefont {Heard}}]{Maartens:1999hf}%
  \BibitemOpen
  \bibfield  {author} {\bibinfo {author} {\bibfnamefont {R.}~\bibnamefont
  {Maartens}}, \bibinfo {author} {\bibfnamefont {D.}~\bibnamefont {Wands}},
  \bibinfo {author} {\bibfnamefont {B.~A.}\ \bibnamefont {Bassett}}, \ and\
  \bibinfo {author} {\bibfnamefont {I.}~\bibnamefont {Heard}},\ }\href
  {\doibase 10.1103/PhysRevD.62.041301} {\bibfield  {journal} {\bibinfo
  {journal} {Phys. Rev. D}\ }\textbf {\bibinfo {volume} {62}},\ \bibinfo
  {pages} {041301} (\bibinfo {year} {2000})},\ \Eprint
  {http://arxiv.org/abs/hep-ph/9912464} {arXiv:hep-ph/9912464} \BibitemShut
  {NoStop}%
\bibitem [{\citenamefont {Barros}\ and\ \citenamefont
  {Nunes}(2016)}]{Barros:2015evi}%
  \BibitemOpen
  \bibfield  {author} {\bibinfo {author} {\bibfnamefont {B.~J.}\ \bibnamefont
  {Barros}}\ and\ \bibinfo {author} {\bibfnamefont {N.~J.}\ \bibnamefont
  {Nunes}},\ }\href {\doibase 10.1103/PhysRevD.93.043512} {\bibfield  {journal}
  {\bibinfo  {journal} {Phys. Rev. D}\ }\textbf {\bibinfo {volume} {93}},\
  \bibinfo {pages} {043512} (\bibinfo {year} {2016})},\ \Eprint
  {http://arxiv.org/abs/1511.07856} {arXiv:1511.07856 [astro-ph.CO]}
  \BibitemShut {NoStop}%
\end{thebibliography}%

\end{document}